# Magnetoelectricity coupled exchange bias in BaMnF$_4$

Shuang Zhou[1], Ji Wang[2,3], Xiaofeng Chang[2,4], Shuangbao Wang[2,4], Bin Qian[5], Zhida Han[5], Qingyu Xu[1,2,*], Jun Du[2,3,*], Peng Wang[2,4], and Shuai Dong[1,*]

[1] Department of Physics, Southeast University, Nanjing 211189, & Key Laboratory of MEMS of the Ministry of Education, Southeast University, Nanjing 210096, & Collaborative Innovation Center of Suzhou Nano Science and Technology, Soochow University, Suzhou 215123, China

[2] National Laboratory of Solid State Microstructures, Nanjing University, Nanjing 210093, China

[3] Department of Physics and Collaborative Innovation Center of Advanced Microstructures, Nanjing University, Nanjing, 210093, China

[4] College of Engineering and Applied Sciences and Collaborative Innovation Center of Advanced Microstructures, Nanjing University, Nanjing 210093, China

[5] Jiangsu Laboratory of Advanced Functional Materials, Changshu Institute of Technology, Changshu 215500, China

## Abstract

Multiferroic BaMnF$_4$ powder were prepared by hydrothermal method. Hysteretic field dependent magnetization curve at 5 K confirms the weak ferromagnetism aroused from the canted antiferromagnetic spins by magnetoelectric coupling. The blocking temperature of 65 K for exchange bias coincides well with the peak at 65 K in the zero-field cooled temperature-dependent magnetization curve, which has been assigned to the onset temperature of two-dimensional antiferromagnetism. An upturn kink of exchange field and coercivity with decreasing temperature was observed from 40 K to 20 K, which is consistent with the two-dimensional to three-dimensional antiferromagnetic transition at Néel temperature (~ 26 K). In contrast to the conventional mechanism of magnetization pinned by interfacial exchange coupling in multiphases, the exchange bias in BaMnF$_4$ is argued to be a bulk effect in single phase, due to the magnetization pinned by the polarization through magnetoelectric coupling.

* Corresponding authors: xuqingyu@seu.edu.cn; jdu@nju.edu.cn; sdong@seu.edu.cn

# Introduction

Multiferroic materials which possess simultaneous magnetic and electric orderings have attracted vast amount of interests, due to their importance in both fundamental physics and practical applications [1-3]. In the past decades, most research attentions on multiferroics have been paid on various transition oxides. Despite the great progress of multiferroic oxides, there are still many drawbacks of these materials, such as low working temperatures and weak polarization/magnetization, which prevents their immediate applications. In fact, there are many multiferroics beyond oxides, e.g. some fluorides [4], which may own better performance and novel physics but have not been well studied.

The fluoride $BaMnF_4$ belongs to a group of isostructural compounds $BaMF_4$ ($M$ = Mn, Fe, Co, Ni, etc.) which have unusual magnetic and electrical properties. Interest in these materials was stimulated by the discovery of both antiferromagnetism and piezoelectricity in 1968 [5]. As a typical multiferroic material, $BaMnF_4$ was sporadically studied in the past decades [6-11], because high quality samples were not easy to be synthesized. The room-temperature structure of $BaMnF_4$ is orthorhombic with space group of $A2_1am$ ($a$ = 5.9845 Å, $b$ = 15.098 Å, $c$ = 4.2216 Å) [9]. $MnF_6$ octahedra linked at the corners form puckered sheets perpendicular to the $b$ axis, which are separated by the intercalation layers of Ba ions, as sketched in Fig. 1(a) and (b). A structural phase transition occurring at $T_C$ = 247 K has been reported, with incommensurate superlattice peaks below $T_C$ at $\vec{q}_1 = (0.392, 0.5, 0.5)$ and its multiples: $\vec{q}_2 = 2\vec{q}_1$ and $\vec{q}_3 = 3\vec{q}_1$ [6]. These wave vectors do not change with decreasing temperature, and such a modulation of structure leads to the distortion of F octahedra [6]. $BaMnF_4$ has a spontaneous polarization $P_S$ = 11.5 μC/cm² along $a$ axis at room temperature [7]. However, the reversal of polarization by electric field has not been observed because of its high conductivity and large potential barrier for the ferroelectric reversal. Thus BaMnF$_4$ is a pyroelectric material and the temperature dependent dielectric constant measurements indicate that the ferroelectric Curie temperature $T_{CFE}$ should be higher than its melting point [6]. With decreasing temperature to around 50 K, a broad peak of magnetization was observed, which was attributed to the two-dimensional antiferromagnetic (2D AFM) transition [6, 12]. The Néel temperature ($T_N$) of BaMnF$_4$ is about 26 K, below which three-dimensional (3D) AFM correlation is established [6]. The ordered magnetic moment was reported to be 4.8 $\mu_B$/Mn$^{2+}$ and lies in the *bc* plane, 9° from *b* axis (Fig. 1(c)), with AFM ordering along both *b* and *c* directions [6]. Weak ferromagnetism with net magnetization along *c* axis induced by magnetoelectric coupling (ME) effect was theoretically predicted [13], and has been experimentally found (0.01 $\mu_B$/Mn) by AFM resonance [14]. However, some other experimental studies, e.g. susceptibility measurement and neutron diffraction, cannot confirm such ferromagnetism [4, 15]. Therefore, the ME effect induced ferromagnetism, or its net magnetization, needs further experimental verification.

Among various magnetic properties, exchange bias (EB) is a very interesting one which has important applications in spintronics [16, 17]. EB was reported mostly in heterostructure systems, such as ferromagnetic (FM)/AFM bilayers, AFM nano-materials with AFM core and spin glass like shell, etc., and the mechanism was generally attributed to interfacial exchange coupling [18-20]. Based on the basic idea of EB in FM/AFM bilayers, such a phenomenon might also be observed in other systems if the FM moment could be pinned through coupling with other physical quantities. Recently, Dong *et al.* reported that EB might be induced by polarization through the ME effect [21]. BaMnF$_4$ has a spontaneous polarization with ME induced AFM spin canting, thus the weak FM moment, if indeed exist, should be coupled to the polarization [22]. Thus, it is reasonable to suspect the EB in BaMnF$_4$ as a fingerprint of ME induced weak ferromagnetism, which has not been reported till now. In this letter, BaMnF$_4$ powder were prepared by hydrothermal method, magnetic transitions to 2D AFM structure at 65 K and 3D AFM structure at 26 K were observed. EB has been observed below 65 K, which has been ascribed to the polarization pinned magnetization through ME effect.

**Results**

Figure 2(a) presents the XRD pattern of BaMnF$_4$ powder. It is clear that the diffraction pattern can be well indexed to the orthorhombic structure of standard JPCDS: 21-0077, confirming the space group of A2$_1$am. No impurity phase can be observed from the diffraction pattern. The size of crystalline grains evaluated by Scherrer formula is greater than 1000 nm, beyond its validity limit, suggesting the much large particle size of our BaMnF$_4$ powder. From the SEM image in Fig. 2(b), BaMnF$_4$ powder show the sheet structure, generally with lateral size in the order of 10 μm, although there are also some smaller fragments. This micro-sheet morphology suggests the anisotropic growth under hydrothermal condition, which is related to the anisotropic crystal structure, as sketched in Fig. 1(a) and (b). As aforementioned, these bilayers of corner-connected octahedra are separated by layers of Ba cations, making a quasi-2D structure. Figure 2(c) shows the electron diffraction pattern from one BaMnF$_4$ sheet with electron beam perpendicular to the sheet plane (inset of Fig. 2(c)). The regular rectangle spotty structure of diffraction pattern suggests the single crystalline structure of each sheet. The ratio of lattice plane distance along the two orthogonal directions (1 and 2 shown in Fig.

2(c)) can be calculated from the diffraction pattern to be 1.42, which coincides well with the ratio of lattice constant $a$ and $c$ ($a$:$c$ = 1.42). This suggests that direction 1 is along $a$ axis and 2 along $c$ axis, while the normal of sheet is along $b$ axis. It is naturally supposed that different facet has different growth speed due to the anisotropic cation and anion arrangements, leading to the different adsorption of growth units [23].

The temperature dependent magnetization (*M-T*) under a magnetic field of 200 Oe was measured after zero field cooling (ZFC) and field cooling (FC) processes with a cooling field ($H_{cool}$) of 200 Oe. As shown in Fig. 3(a), the ZFC *M-T* curve shows a clear maximum at around 50 K, while continuous increase of magnetization with decreasing temperature was observed in the FC *M-T* curve. This peak temperature agrees well with the previous report of a broad peak at around 50 K in the *M-T* curve, which was ascribed to the 2D AFM transition [12]. The previous magnetic investigations were mainly performed on single crystalline samples [15, 24]. Isotropic magnetic properties were observed at *T*>30 K, below which a continuous decrease of magnetization was observed with a field along *b* axis, while a minimum was observed at around 25 K, and then increased with further decreasing temperature with a field perpendicular to *b* axis. Due to the powder nature of our samples, the ZFC *M-T* curve is the average result with a field along random crystalline orientations. Thus, although the minimum of ZFC magnetization can still be observed, its temperature is much lower than 25 K, and the increase of magnetization with further decreasing temperature is much weaker. A clear kink can be observed at around 26 K in the FC *M-T* curve, which coincides with the reported $T_N$ of 3D AFM transition [6].

A splitting of ZFC and FC *M-T* curves is observed below 65 K, which is assigned to $T_{irr}$

(irreversible temperature) here and has not been reported in BaMnF$_4$ previously. Such a splitting is generally observed in those systems with unstable magnetic structures, such as spin glass, superparamagnetism, etc. [25-27]. The typical character of spin glass is the memory effect [25-26]. Memory effect experiment was performed following the reported protocol [28]. A ZFC *M-T* curve was measured with a stop at 32 K for 6 hours during the cooling process first, and then another ZFC *M-T* curve was measured with continuous decrease of temperature. No signature of spin glass was observed [29]. We further performed the AC magnetic susceptibility measurements on BaMnF$_4$ powder with various frequencies, and the results are shown in Fig. 4. A broad peak can be observed for all susceptibility (real and imaginary) curves, and there's no frequency-dependent peak shift, which can also exclude the possible spin glass contribution. The weak ferromagnetism has been reported to be induced by the ME effect in 3D antiferromagnetism below $T_N$ [13]. The onset magnetization in FC *M-T* curve below 65 K suggests that weak ferromagnetism might be induced by polarization through the ME effect even in the 2D AFM phase. We attributed 65 K to the onset temperature of the 2D AFM structure. The observed irreversibility can be understood by the magnetic anisotropy of the ME induced spin-canted magnetic domains [27]. There might be metastable states below 65 K, due to the competition between magnetic crystalline anisotropy, and domain wall pinning. The energy configuration is determined by applied field and temperature, influencing the spin flipping from metastable states to stable states. Similar to increasing temperature, increasing the applied field can also decrease the energy barrier and suppress the irreversibility [27]. The ZFC and FC *M-T* curves under magnetic field of 50 kOe were further measured, as shown in Fig. 3(b). The nearly overlapping of ZFC and FC *M-T* curves indicates that the Zeeman energy of weak ferromagnetic moment under field of 50 kOe is enough to overwhelm the energy barrier. The broad peak at around 50 K can still be observed for both curves. Furthermore, clear minima at around 28 K corresponding to the 3D AFM transition, can be clearly observed for both ZFC and FC *M-T* curves. It has been reported that the spins of Mn ions are easily aligned to the field direction in two-dimensional AFM phases [15], suggesting that the magnetic field should be smaller to observe the low-temperature anomaly. With decreasing field to 50 Oe during the ZFC *M-T* measurement (inset of Fig. 3(a)), a small peak at 65 K can be clearly observed, confirming an AFM transition at $T_{irr}$.

Figure 5(a) shows the *M-H* loops of BaMnF$_4$ measured at 300 K and 5 K after cooling from 300 K with $H_{cool}$ of 200 Oe. The loop of 300 K is a straight line, revealing that BaMnF$_4$ is paramagnetic at room temperature. At 5 K, a clear hysteretic *M-H* loop has been observed, with remanent magnetization of 0.015 emu/g (7.2×10$^{-4}$ $\mu_B$/Mn$^{2+}$), suggesting the weak ferromagnetism. Theoretically, it has been predicted that there is a tiny FM component along *c* axis: $M_c \sim 4\pi\alpha_{ac}P_a$, where $P_a$ is the ferroelectric polarization, and $\alpha_{ac}$ the magnetoelectric tensor element related to the ordering of antiferromagnetism [22]. However, such ME induced weak ferromagnetism was doubted recently, since no FM component was observed using neutron diffraction and susceptibility measurements [10, 15]. The exact origin for this discrepancy is not clear at this moment. The observed hysteretic *M-H* loop in our samples at 5 K confirms the ME induced weak ferromagnetism. However, its remanant magnetization is much weaker than the reported value of 0.01 $\mu_B$/Mn, which may be the reason for the discrepancy between different experimental methods. In contrast to ordinary *M-H* loops, an upward deviation from straight line can be observed with increasing magnetic field in the 5 K *M-H* loop, which has been ascribed to the spin-flop transition at low temperatures [12]. For the *M-H* loop at 5 K with $H_{cool}$ of 200 Oe, a clear shift to left has been observed (left inset of Fig. 5(a)), indicating the EB, which has not been observed in the loop at 300 K. To exclude the possible measuring errors, *M-H* loop at 5 K with $H_{cool}$ of -200 Oe was measured and is shown in the right inset of Fig. 5(a). The clear right shift confirms the observed EB.

It has been reported that if the applied maximum field is not enough to reverse the magnetization of FM phase, the minor loop might also lead to the similar phenomenon of EB [30-33]. As the coercivity of our sample is only about 100 Oe, the applied field of 10 kOe is much larger than the coercivity, the *M-H* loops measured under maximum field of 10 kOe are not likely to be minor loop. The system can be considered effectively saturated if the ascending and descending branches of its hysteresis loop coincide for fields higher than the anisotropy field [34]. The inset of Fig. 5(b) shows that the magnetization is reversible in high fields, the arrow marking the field of loop closure ($H_{irr}$ ~ 7.6 kOe). The applied field of 10 kOe is much larger than $H_{irr}$, which excludes the minor loop effect. Furthermore, if the observed EB is due to minor loop, the *M-H* loop measured with maximum field of 10 kOe at 5 K should shift to left in the case of $H_{cool}$ of -200 Oe, in contrast to the experimental observed positive shift. We applied different maximum field up to 70 kOe to measure the *M-H* loops, and the results are shown in Fig. 5(c). As can be seen, with increasing maximum field to above 10 kOe, the descending parts of *M-H* loops are nearly overlapping, while significant shift to right can be observed in the ascending parts. The inset shows the $H_E$ and $H_C$ in dependence on the maximum field. $H_E$ and $H_C$ are defined as $H_E$=-($H_L$+$H_R$)/2 and $H_C=(H_R-H_L)/2$, where $H_L$ and $H_R$ are the left and right coercivities, respectively. $H_E$ decreases with increasing the maximum field and starts to saturate with higher field. $H_E$ of 24 Oe can still be observed with maximum field of 70 kOe, which confirms the intrinsic nature of EB. The continuous decrease of $H_E$ is mainly due to continuous right shift of the ascending parts of *M-H* loops. This might be understood by that the reversed magnetization might be pinned in metastable states with increasing the reversed maximum field, which will become harder to be reversed with positive field [35].

In order to explore the relationship between temperature and EB, we measured the *M-H* loops at various temperatures after cooling under field of 200 Oe. Figure 5(b) shows $H_E$ and $H_C$ as a function of temperature. $H_C$ was also measured at 5 K under ZFC process, which is about 138 Oe and slightly larger than the value measured under FC process (star in Fig. 5(b)). With temperature above $T_{irr}$=65 K, both $H_E$ and $H_C$ are nearly zero. As $H_C$ is closely related to the remanent magnetization due to weak ferromagnetism, the near zero $H_C$ suggests the near zero remanent magnetization. Thus, above $T_{irr}$=65 K, there is no ME induced weak ferromagnetism. Generally, the temperature where EB disappears is called blocking temperature $T_B$. The coincidence of $T_B$ and disappearing temperature of remanent magnetization suggests the coupling between weak ferromagnetism and EB. With decreasing temperature, both $H_E$ and $H_C$ show the tendency of saturation. With further decreasing temperature below 40 K, an upturn of both $H_E$ and $H_C$ can be observed, which again start to saturate below 20 K. This upturn of $H_E$ and $H_C$ can be understood by the emergence of 3D AFM structure.

**Discussion**

EB in complex bulk oxides generally results from interfaces between ferromagnetic, antiferromagnetic, or spin glass regions [29]. For those antiferromagnetic oxides showing ferromagnetic-like or spin glass surface, previous experiments were mostly focused the nanosized systems, in which the similar behavior (remnant $M$ and EB) were indeed observed [36-39]. However, the grain size of our power sample is in the micrometer scale, thus the surface ratio is much reduced comparing with those nanosized materials. As has been discussed, the magnetic contribution of spin glass has been excluded in our $BaMnF_4$ powders from the memory effect and AC magnetic susceptibility measurements. The magnetic contribution from surface disorders can be safely excluded.Thus the possible mechanism of exchange coupling between AFM core and spin glass like surface can be excluded [10, 18, 40-43]. Furthermore, the ME induced weak ferromagnetic moment in the particle core might also be pinned by the disordered surface spins [44, 45]. However, since this is an interface effect, the EB will decrease drastically with increasing the thickness of FM layer, similar to the conventional FM/AFM bilayers [46]. The influence of $H_{cool}$ on $H_E$ and $H_C$ was further studied, as shown in Fig. 6. The sample was cooled down from 300 K to 5 K under different field up to 50 kOe. We notice that there are two regions for the variation of $H_E$ and $H_C$. At low cooling field up to around 5 kOe, $H_E$ increases abruptly while $H_C$ decreases abruptly with increasing $H_{cool}$. At high cooling field, both $H_E$ and $H_C$ show saturated behavior. In conventional AFM/FM bilayer system, both $H_E$ and $H_C$ increase with increasing $H_{cool}$ in the low field range [47, 48], while in the exchange bias system with frustrated AFM spins at interface, $H_E$ tends to decrease [49], even changes the sign with increasing $H_{cool}$ in the low field range [50-52]. The increase of $H_E$ with increasing $H_{cool}$ in the low field region further excludes the possible coupling with the surface disorder spins. When the cooling field is low, it is not enough to make all the weak ferromagnetic magnetization aligned. With the increase of cooling field, the degree of the weak ferromagnetic moment alignment is enhanced, which reduce the effect of the averaging of anisotropy due to randomness [53]. Interestingly, $H_C$ shows a decrease with increasing $H_{cool}$ in the low field range, which is rather unusual and cannot be explained at this moment.

Instead, we'd like to attribute the origin of EB to its magnetoelectricity, which can give a self-consistent scenario to understand above experimental observations. First, the critical temperature for the 2D AFM transition is $T_{irr}=T_B=65$ K, instead of previous supposed 50 K. In addition, Scott also mentioned that in BaMnF$_4$ the spins order in planes up to until approximately $T=3T_N$ [4], which is closer to our observed blocking temperature of 65 K rather than the reported 50 K. Below this temperature, weak FM moments can be induced by the polarization and AFM ordering through ME effect in BaMnF$_4$ [22], as sketched in Fig. 7. Since the ferroelectric Curie temperature $T_{CFE}$ of BaMnF$_4$ is very high (higher than its melting point), $P_a$ should be robust during our magnetic measurements with temperature decreasing from 300 K to 5 K. Due to the ME effect, the easy axes of neighboring Mn ions are not exactly in line, but with a small angle. The spin structures after FC process are shown in Fig. 7(a) and (b) with $H_{cool}$ of opposite direction. The induced $M_c$ due to spin canting will be aligned along the direction of $H_{cool}$, even when there's only 2D AFM correlation below 65 K. Of course, such an ME effect also works after the 3D AFM transition at 26 K. After the establishment of such spin structures at low temperatures, the magnetization has preferred orientation along the direction of $H_{cool}$. With applied field ($H_{appl}$) opposite to the direction of $H_{cool}$ during the *M-H* loop measurements, the spins $S_1$ and $S_2$ slightly rotate away from the easy axes with net magnetization along the direction of $H_{appl}$, as shown in Fig. 7(c) and (d) ((a) to (c) and (b) to (d)). However, the spins $S_1$ and $S_2$ will rotate back to their easy axes after switching off the field. The spin structures tend to switch back to their initial states ((c) to (a) and (d) to (b)), respectively. Thus, $H_L$ is larger than $H_R$ in the case of positive $H_{cool}$ and $H_L$ is smaller than $H_R$ in the case of negative $H_{cool}$, leading to the observe EB. In contrast to the general mechanism of interfacial exchange coupling in heterostructures, here the EB in BaMnF$_4$ is a bulk effect, due to the magnetization pinned by polarization through ME effect. EB is a special method to verify the ME induced weak ferromagnetism in multiferroics. Following this mechanism, the electrical-controllable EB is expected by flipping the polarization. Although there are some experimental difficulties for BaMnF$_4$ studied here, the ferroelectric polarization flip may be possible in near future, e.g. using ion substitution to reduce its ferroelectric energy barrier and suppress its leakage.

In summary, we have successfully synthesized pure BaMnF$_4$ powders by hydrothermal method. The powders are single crystals in sheet structure, with *b* axis as the normal of sheet plane. Weak ferromagnetism induced by the ME effect has been confirmed in both 2D and 3D AFM phases by the observation of hysteretic *M-H* loops. A significant EB after FC treatments was observed, which is strongly dependent on the temperature. The EB is visible only below 65 K, which has been assigned to the temperature of 2D AFM transition. An upturn was observed in the temperature dependent $H_E$ and $H_C$ from 40 K to 20 K, coinciding with $T_N$ of 26 K, where the 3D AFM transition happens. The mechanism of EB has been explained by the magnetization pinned by polarization through ME effect. In a word, the magnetic poling can fix the sign of $\alpha$ without the help of electric field. This would be a very unique property of ME antiferromagnets. More direct measurements, e.g. magnetoelectric measurements after magnetic field cooling processes, can further check the sign of ME tensor component. Following this mechanism, the electrical-controllable EB might be realized and applied in spintronics.

**Methods**

BaMnF$_4$ powders were synthesized by hydrothermal method [54]. Appropriate amounts of BaF$_2$ and Mn(CH$_3$COO)$_2$·4H$_2$O were dissolved in trifluoroacetic acid solution (5 ml CF$_3$COOH and 10 ml distilled water), and the diluted solution was formed by magnetic stirring, which was then put into an autoclave. The autoclave was gradually heated to 220 °C, held for 20 hours, then slowly cooled down to room temperature. The upper remaining liquid was discarded, and the precipitates were kept and washed with ethanol for several times. The washed products were placed in a vacuum drying oven and dried at 95 °C, resulting in the final pale pink powder.

The structure of samples was studied by X-ray diffraction (XRD, Rigaku Smartlab3) using a Cu Kα radiation and transmission electron microscope (TEM, Tecnai F20). The morphology was studied by a scanning electron microscope (SEM, FEI Inspection F50). The DC magnetization was measured by a superconducting quantum interference device (SQUID, Quantum Design) from 5 K to 300 K. AC magnetic measurements were carried out using a physical property measurement system (PPMS, Quantum Design).

## Acknowledgments

This work is supported by the State Key Programme for Basic Research of China (2014CB921101), the National Natural Science Foundation of China (51172044, 51471085, 51322206), the Natural Science Foundation of Jisangsu Province of China (BK20151400).

## Author contributions statement

Q.X. conceived and designed the research. S.Z. carried out the experiment. S.D. developed the theoretical explanation, X.C., S.W. and P.W. did the TEM characterization. J.D. and J.W. carried out the magnetic measurements. B.Q and Z.H. measured the AC susceptibility. S.Z. Q.X., and S.D. wrote the paper.

## Additional information

Competing financial interests: The authors declare no competing financial interests.

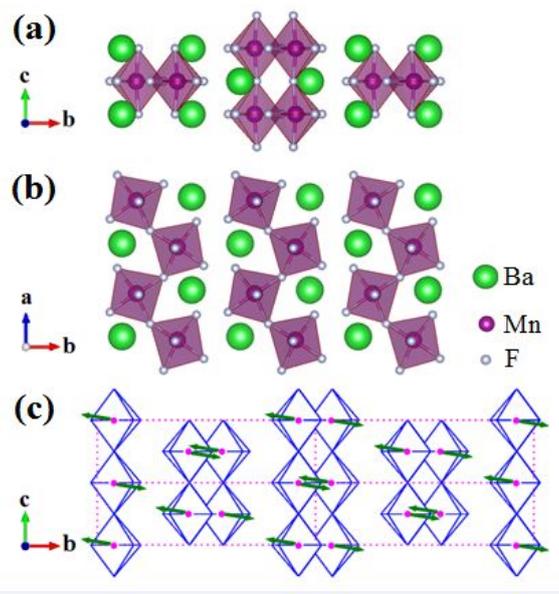

**Figure 1.** The schematic crystal structure of BaMnF$_4$ in (a) *bc* plane and (b) *ab* plane. (c) the AFM structure, where the AFM axis lies in *bc* plane and is directed about 9° from *b* axis.

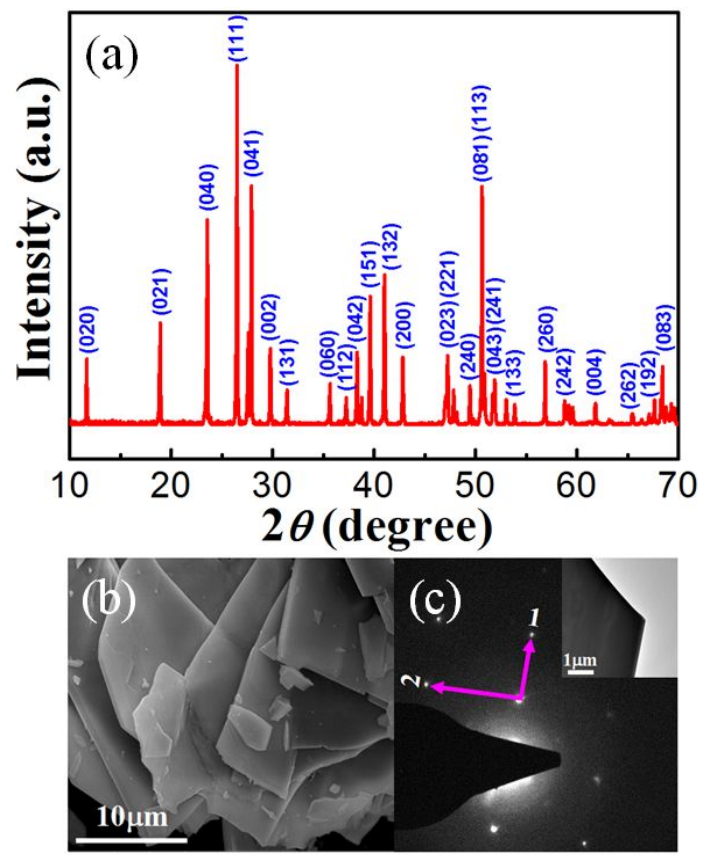

**Figure 2.** (a) The XRD pattern and (b) SEM image of BaMnF$_4$ powders. (c) The electron diffraction pattern from one BaMnF$_4$ sheet with electron beam perpendicular to the sheet, as shown in the inset.

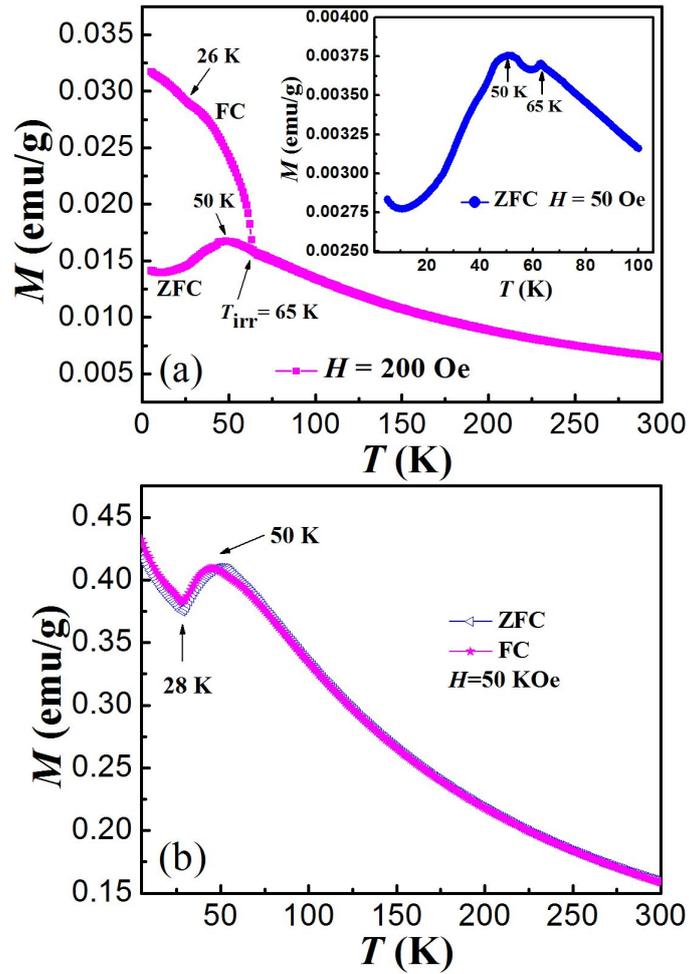

**Figure 3.** (a) ZFC and FC *M-T* curves measured under *H* = 200 Oe (main plot). ZFC *M-T* curve measured under *H* = 50 Oe (inset). (b) ZFC and FC *M-T* curves measured under *H* = 50 kOe.

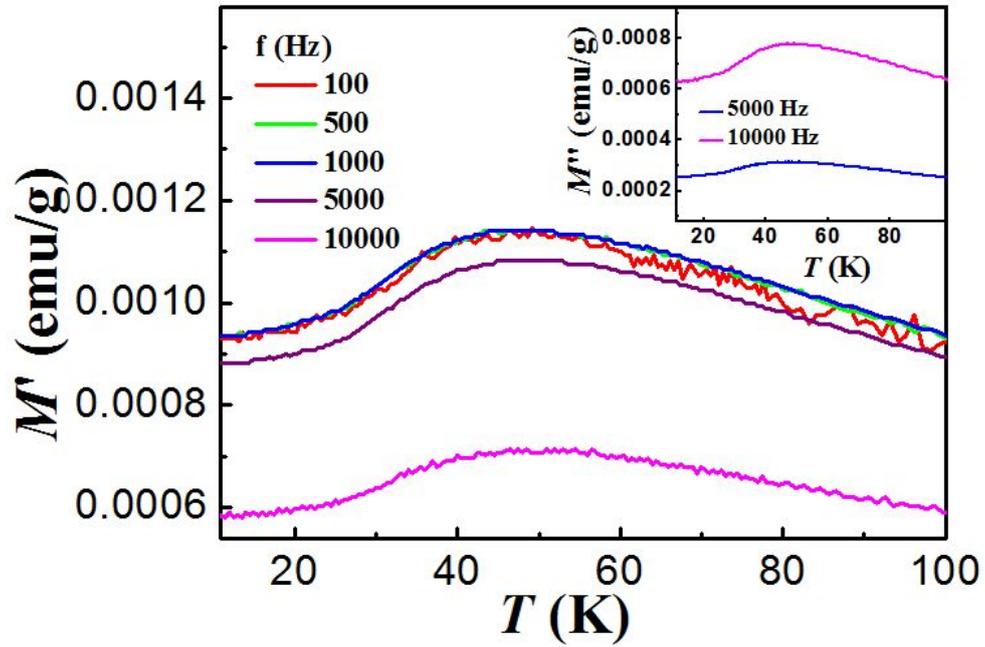

**Figure 4.** AC magnetic susceptibility *M'* (real part) of BaMnF$_4$ measured under various frequencies. Inset: the imaginary part *M''*. It should be noted that the value of *M''* is too small to be measured under low frequencies.

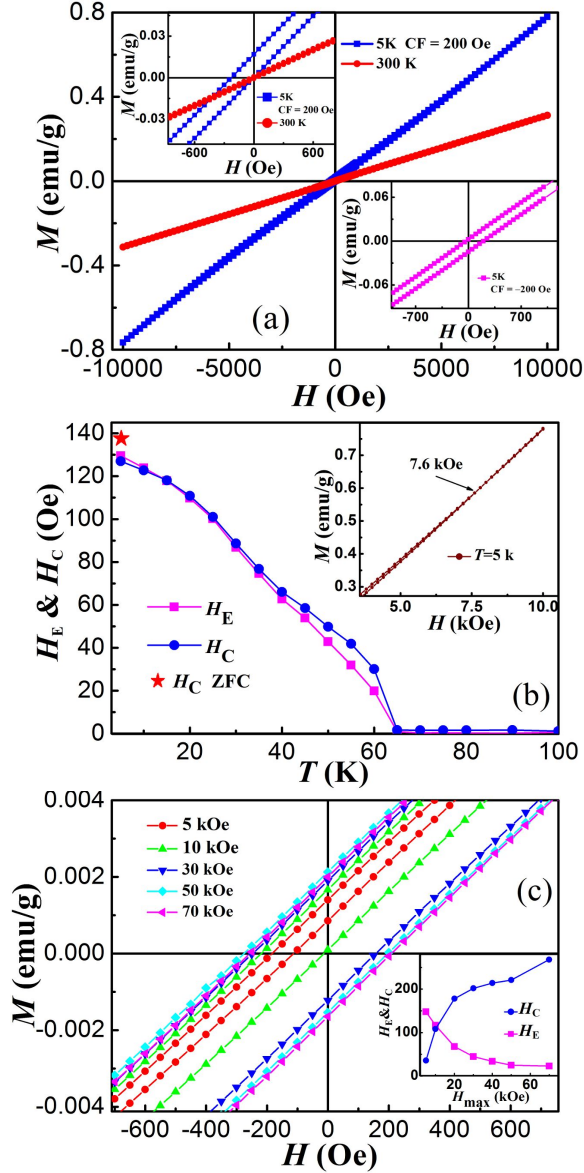

**Figure 5.** (a) *M-H* loops of BaMnF$_4$ at *T* = 300 K and *T* = 5 K after cooling with a field of 200 Oe with the enlarged view in left inset. Right inset: the enlarged view of *M-H* loop measured with cooling field of -200 Oe. (b) Temperature dependence of $H_E$ and $H_C$. The inset shows the enlarge view of *M-H* loop measured at 5 K with maximum field ($H_{max}$) of 10 kOe after cooling under field of 200 Oe. The arrow indicates the closure of hysteresis loop at $H_{irr}$. (c) Enlarged view of *M-H* loops measured under different $H_{max}$ at 5 K after cooling under field of 200 Oe. The inset shows the dependence of $H_E$ and $H_C$ on $H_{max}$.

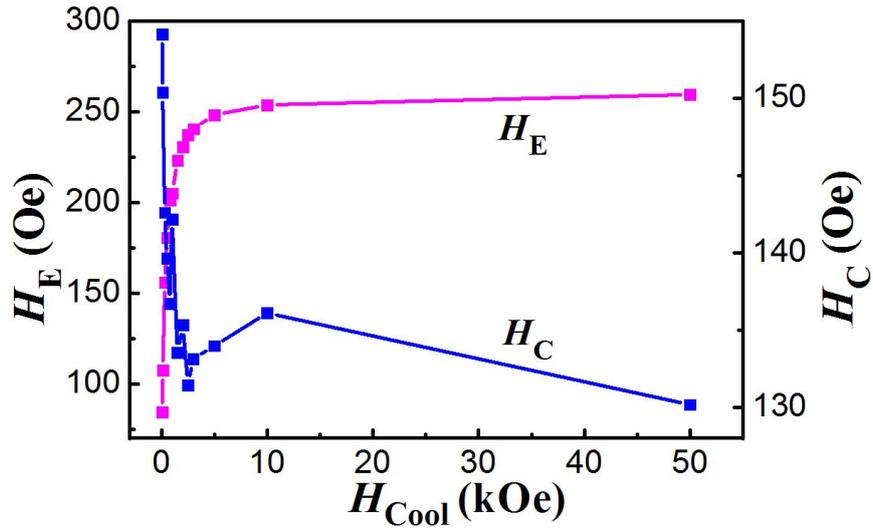

**Figure 6.** The dependence of $H_E$ and $H_C$ on $H_{cool}$, measured at 5 K with maximum field of 10 kOe.

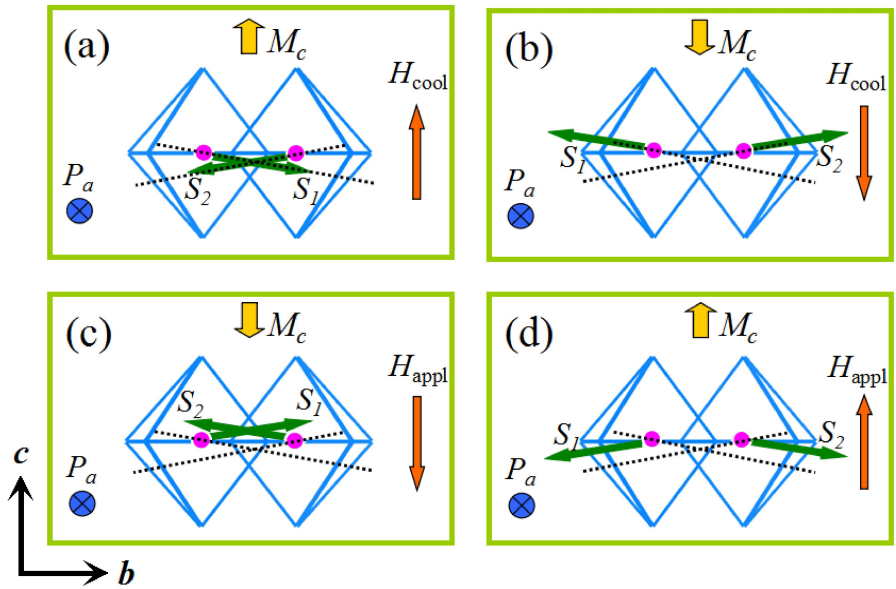

**Figure 7.** The illustration of magnetoelectricity induced EB. With given polarization $P_a$, the canting of neighboring AFM spins $S_1$ and $S_2$ can be affected by the cooling field direction due to Zeeman energy of net moment $M_c$. The dashed lines are the easy axes of neighboring Mn ions.